\documentstyle[prd,aps,epsf]{revtex}

\def\be{\begin{eqnarray}}
\def\en{\end{eqnarray}}

\begin{document}

\title{Electromagnetic Generators and Detectors of Gravitational Waves
\footnote{Invited talk at the first conference on High-Frequency
Gravitational Waves, May 2003, The MITRE Corporation, McLean, Virginia, USA}}
\author{L. P. Grishchuk}
\address{Department of Physics and Astronomy, Cardiff University, 
Cardiff CF24 3YB, United Kingdom \\
and  Sternberg Astronomical Institute, Moscow University, Moscow
119899, Russia\\ {\rm E-mail: grishchuk@astro.cf.ac.uk}}
\maketitle

\begin{abstract}
The renewed serious interest to possible practical applications of 
gravitational waves is encouraging. Building on previous work, I am arguing 
that the strong variable electromagnetic fields are appropriate systems for the
generation and detection of high-frequency gravitational waves (HFGW). The 
advantages of electromagnetic systems are clearly seen in the 
proposed complete laboratory experiment, where one has to ensure the efficiency
of, both, the process of generation and the process of detection of HFGW. 
Within the family of electromagnetic systems, one still has a great variety of 
possible geometrical configurations, classical and quantum states of the 
electromagnetic field, detection strategies, etc.  According to evaluations
performed 30 years ago, the gap between the HFGW laboratory signal and its 
level of detectability is at least 4 orders of magnitude. Hopefully, new 
technologies of today can remove this gap and can make the laboratory 
experiment feasible. The laboratory experiment is bound to be expensive, but 
one should remember that a part of the cost is likely to be reimbursed from the
Nobel prize money ! Electromagnetic systems seem also appropriate for the 
detection of high-frequency end of the spectrum of relic gravitational waves. 
Although the current effort to observe the stochastic background of relic 
gravitational waves is focused on the opposite, very low-frequency, end of the 
spectrum, it would be extremely valuable for fundamental science to detect, or 
put sensible upper limits on, the high-frequency relic gravitational waves. 
I will briefly discuss the origin of relic gravitational waves, the expected 
level of their high-frequency signal, and the existing estimates of its 
detectability.
\end{abstract}

\section{Introduction}

The engineers and enterpreneures are rightly seeing in 
gravitational waves (g.w.) a new opportunity, even though, 
at this moment of time, this opportunity does 
not look like having a chance of becoming practical in foreseeable future. 
However, we have many examples of wrong prognoses. It is known that when the 
discoverer of the nucleus, Lord Rutherford, was asked about the possible 
practical applications of nuclear energy, he answered ``never". It is very 
likely that the gravitational radiation will eventually become useful, 
say, for purposes of communication. But in any case, it is difficult 
to imagine that we will go along without attempting to
demonstrate in laboratory conditions the possibility of 
controllable generation and detection of gravitational waves.
The laboratory experiment is a necessary first step to possible 
practical applications of g.w., and it is this
first step that will be mostly discussed in this contribution.  
 
Gravitational waves are less mysterious than one can gather from popular
accounts. The relativistic Einstein's gravity is usually described in 
geometrical terms, as curvature of space-time. Correspondingly, 
gravitational waves are often described 
as ``oscillations of space-time itself". This characterisation is 
quite puzzling and distressing for a practically-minded person. 
It makes defunct the usual physical intuition 
based on electromagnetic waves and other propagating physical
fields. It makes one to suspect that there may be something artificial and 
unreal in the ``waves of space-time itself". But this characterisation is 
only a matter of unfortunate language. Geometrical concepts are useful for 
some purposes, but they are greatly misleading for others. 
One should remember that although the Einstein gravity can be 
described in geometrical terms, it is also a normal physical field. 
The gravitational field is universal in all its interactions with 
itself and other fields and matter, and this is what 
makes the geometrical formulation of relativistic gravity possible, but 
the geometrical formulation is not necessary and is not obligatory. 
In fact, it would be a nightmare to try to discuss the laboratory 
gravitational-wave experiment in terms of differential geometry and
``oscillating space-time", rather 
than in the engineering terms of the emitted and absorbed physical 
radiation. 

The relativistic gravitational field is governed by the non-linear 
wave equations - the Einstein equations. However, in
many situations one can neglect the non-linearities of the
gravitational field. In particular,
very often one can neglect the interaction of gravitational waves
with other gravitational fields and with themselves. In this case, 
we come to the notion of linearized gravity and weak gravitational
waves. Certainly, in any laboratory experiment we will be dealing
with extremely weak gravitational waves. Weak g.w. are very far 
from being mysterious; they may even appear quite boring and 
disappointing. Indeed, there is not so much difference with electromagnetic 
waves in terms of their general properties, but gravitational waves 
interact with matter and other fields much-much less effectively 
than electromagnetic waves. Gravitational waves carry their energy practically
without scattering and absorption. This is why it is so difficult to
detect astronomical gravitational waves. But the other side 
of this difficulty is the tremendous penetrating capability of gravitational 
waves. This is why they are so important as a tool of astronomical research,
and this is why they attract attention as a possible unique means of 
communication.   

The relativistic gravitational field can be described by 10 components of 
the $4 \times 4$ symmetric dimensionless tensor $h^{\mu \nu}$. The
components of $h^{\mu \nu}$ are functions of 
time and spatial coordinates $x^{\alpha} = (ct, x, y, z)$.
The functions $h^{\mu \nu}$ obey the nonlinear wave-like dynamical equations -
the Einstein equations. The energy-momentum tensor $t^{\mu \nu}$ of the 
gravitational field is calculable from the field tensor $h^{\mu \nu}$. 

The linearised gravitational waves satisfy the wave equation
\begin{equation} 
\label{Lgw}
{h^{\mu\nu ,\alpha}}_{,\alpha} + 
\eta^{\mu\nu}{h^{\alpha\beta}}_{,\alpha ,\beta} - 
{h^{\nu\alpha ,\mu}}_{,\alpha}- {h^{\mu\alpha ,\nu}}_{,\alpha} = 0,
\end{equation} 
where the ordinary derivative is denoted by a comma and 
$\eta^{\mu \nu}$ is the metric tensor of the Minkowski space-time:
\begin{equation}
\label{Mi}
{\rm d} \sigma^2 = \eta_{\mu \nu} {\rm d}x^{\mu} {\rm d} x^{\nu} =
c^2{\rm d}t^2 - {\rm d}x^2 - {\rm d}y^2 - {\rm d}z^2.   
\end{equation} 
The first term in Eq. (\ref{Lgw}) is the familiar d'Alembert (wave) operator: 
\[
{h^{\mu\nu ,\alpha}}_{,\alpha} = \left(\frac{\partial^2}{c^2 \partial t^2}-
\frac{\partial^2}{\partial x^2} -  \frac{\partial^2}{\partial y^2} -  
\frac{\partial^2}{\partial z^2}\right)h^{\mu \nu}.  
\]

A plane-wave solution to Eq.~(\ref{Lgw}) is given by 
\begin{equation} 
\label{plw}
h^{\mu\nu}= a^{\mu\nu} e^{i  k_{\alpha} x^{\alpha}}+ {\it c.c.},  
\end{equation} 
where $k_{\alpha}k^{\alpha} = 0$, reflecting the fact that a g.w. 
propagates with the speed of light. Because of this condition, the field
equations (\ref{Lgw}) require the 10 components of the constant matrix 
$a^{\mu\nu}$ to satisfy 4 constraints: $a^{\mu\nu}k_{\nu} = 0$.
Among the remaining 6 components, only 2 degrees of freedom 
(sometimes called the TT-components $\tilde{a}^{ij}$) are 
physically important, in the sense that it is only these degrees of freedom
that fully determine the observational manifestations of the plane 
wave and its energy-momentum characteristics. Indeed, it is easy to show
that the gravitational energy-momentum tensor 
\begin{equation} 
\label{gwtmn}
t_{\mu\nu} =\frac{c^4}{32 \pi G}\left[{h^{\alpha\beta}}_{, \mu}
h_{\alpha\beta , \nu} - \frac{1}{2} h_{, \mu}h_{, \nu} \right]
\end{equation}  
depends only on the TT-components of the field and its two independent
amplitudes $h_{+}$ and $h_{\times}$:
\begin{equation} 
\label{gwenm}
t_{\mu\nu}= \frac{c^4}{32 \pi G} k_\mu k_\nu \left[2 \tilde{a}^{ij} 
\tilde{a}^{*}_{ij}\right]= 
\frac{c^4}{32 \pi G} k_\mu k_\nu \left[h_{+}^2 + h_{\times}^2 \right]. 
\end{equation} 
In this expression we have dropped (as we normally do in the case of 
energy-momentum tensor for the electromagnetic waves) the purely 
oscillatory terms.

The amplitudes $h_{+}, h_{\times}$ are determined by the source of 
the gravitational waves. To find the amplitudes, we should 
replace the zero in the right hand side of Eq. (\ref{Lgw})
by the source term $(16 \pi G/ c^4) T^{\mu\nu}$, where $T^{\mu\nu}$ is the
energy-momentum tensor of the source, and seek the retarded solutions
to the inhomogeneous wave equation. For example, for a pair of stars 
in a circular orbit, with masses of the stars $M_1$, $M_2$, 
located at the distance $R_0$ from
us, and after averaging over the orbital period and orientation of the 
orbital plane, we obtain  
\be
\label{amplb}
h = \left( \langle h_{+}^2 \rangle + \langle h_{\times}^2\rangle \right)^{1/2}=
\left( \frac{32}{5}\right)^{1/2} \frac{1}{R_0} \frac{G^{5/3}}{c^4}
\frac{M_1 M_2}{(M_1 +M_2)^{1/3}} (\pi f)^{2/3},
\en
where the emitted g.w. frequency $f$ (in $Hz$) is twice the orbital frequency.

Roughly, the characteristic amplitude $h$ from a given source is given by
\be
\label{ampl}
h_{ij} \approx \frac{G}{c^4} \int \frac{T_{ij}(t_{ret})}{r} d V ~~~~~
{\rm and}~~~~~  
h \approx \frac{1}{R_0} \frac{G M}{c^2} \left(\frac{v}{c}\right)^2,
\en
where $M$ is the total mass of the source and $v$ is the characteristic 
velocity of the matter bulk motion. This formula
is quite universal, and it can also be written in terms of the characteristic
variable stresses $\sigma$ within the source, and the source's volume $V$:
\be
\label{ampl2}
h \approx \frac{1}{R_0} \frac{G}{c^4} \sigma V.
\en
As long as the retardation effects inside the
source can be neglected, this formula is equally well applicable to 
the radiating systems of any nature - in cosmos and in laboratory, 
mechanical and electromagnetic. 

\section{Current status of the astronomical program and its comparison
with a HFGW laboratory program}

The dimensionless number $h$ is a very convenient characterisation
of the strenght of a given g.w. and its detectability. Under the action
of a gravitational wave, a pair of free masses, initially separated by the
distance $l$, experience relative oscillatory displacements    
proportional to the incoming wave amplitude $h$: $\delta l/l \approx h$. 
Very powerful astronomical sources, currently under intense experimental
g.w. searches \cite{ligo}, \cite{cutlthorne}, \cite{gr03}, 
produce something like $h \approx 10^{-22}$ at Earth. This is an 
increadibly small number. It enters any conceivable method of detection 
of gravitational waves and explains why it is so difficult to observe 
them. For example, in a $4km$-long laser interferometer, such as LIGO, we 
need to beat all the noises and measure the mirror's displacements at the 
level of $4 \times 10 ^{-17} cm$. At the same time, the flux of energy
$F= ct^{0i}$ at Earth from the discussed sources is quite impressive by 
astronomical standards. At the representative frequency
$f = 200 Hz$, the flux reaches $F \approx 3 \times 10^{-2} erg/sec~ cm^2$.
But most of this energy passes through the g.w. detectors without scattering
and absorption. This is why, in the gravitational-wave physics, 
such characteristics as {\it watts} and {\it joules} may be misleading.
The dimensionless amplitude $h$ is more adequate. But in any case, 
whatever the units in which the analysis is being carried out, only a joint 
discussion of the complete system (emitter plus detector) 
gives the correct evaluation of what is ``small" or ``big", easy to detect
or extremely difficult to detect. 

The recently assembled LIGO interferometers are approaching
their planned level of sensitivity. It appears that, at the time of writing,
there still exists a gap in one order of magnitude, in terms of $h$,
between the actually reached level of sensisitivity and the design 
sensitivity \cite{ligo}. One should clearly understand what is likely to
happen when the design level of sensitivity of the initial interferometers 
is finally achieved. Baring the extremely fortunate surprises, we will 
probably be able to see only the most powerful, but rare, sources, such as 
coalescing binary stellar-mass black holes. Less powerful, even if more
numerous, sources - coalescing binary neutron stars, are unlikely to be
seen, because the expected signal-to-noise ratio is somewhat smaller than 1. 
For a guaranteed detection of astronomical sources, the experimenters
will have to improve the sensitivity by one further order of magnitude, 
as compared
with the design sensitivity of the initial instruments. This is the goal
of the so-called advanced interferometers, and this goal will probably
be reached by 2007. (For more details on the current status of the 
gravitational wave astronomy see, for example, 
\cite{cutlthorne}, \cite{gr03}.) 

There is little doubt that the astronomical
program will completely dominate, and rightly so, any other experimental
efforts in the gravitational-wave physics for some time to come. Having 
said that, one can still wonder whether a laboratory experiment is much 
more difficult to realise than to build equipment for the observation of
cosmic gravitational waves. Of course, the justifications for these efforts
are totally different. In gravitational-wave astronomy we directly explore 
the fascinating Universe, whereas in the laboratory experiment we are
likely to confirm a theory (true, very fundamental theory, but anyway 
tested also by other means), with very remote prospects for practical 
applications of gravitational radiation. However, the justification
for the laboratory experiment is sufficiently convincing. 
A more difficult question is its feasibility. Here, one will have to 
admit that the both enterprises are very difficult. Surprisingly, the 
laboratory program does not appear to be 
unacceptably more difficult than the cosmic program. If we take as a 
benchmark the 4 orders of magnitude separating the detecting and
generating capabilities in laboratory (see Sec. IV below), it is like having 
1 dollar instead of required 10000. But, strictly speaking, in the 
cosmic program we also have, at the time of writing, only 
something like 1 dollar instead of required 100 (see above).   
The difference between the two programs is substantial, but not 
ridiculously large.

As an additional motivation for the experimental work on HFGW, there
are arguments showing that it can be complementary and useful for the 
astronomical program. First, there exists the fundamentally important 
cosmological
signal - relic gravitational waves. The high-frequency end of the
spectrum is the most natural interval for the high-frequency techniques,
and first of all for electromagnetic detectors (see Sec V below). 
Second, part of the high-frequency studies with electromagnetic 
detectors may eventually be useful for laser interferometers of the 
next generation. Indeed, it is quite likely that in order to reach 
the required level of sensitivity we will be forced to implement 
the sophisticated techniques such as squeezed light and 
quantum-nondemolition measurements (see, for
example, \cite{brthorne}) and this is where the 
expertise of the HFGW community can be useful.  

\section{Efficiency of gravitational-wave emitters}

Astronomical sources are immensely  more powerful than any conceivable
sources at Earth, but they are very far away from us, and not under 
our control. In contrast, in laboratory, one can place the emitter 
and the detector very close to each other, choose the appropriate 
materials, implement the coherence and focusing of the
emitted radiation, use the advantages of the resonant detection, etc. 
One usually illustrates the hopelessness of laboratory
g.w. experiments by giving an example of the meager gravitational radiation
generated by a rotating massive rod. Well, if you are so naive 
that you plan to rotate a rod, then the enterprise may indeed
be hopeless. But surely there must exist something smarter. Let us give
a general comparative analysis of possible mechanical and
electromagnetic systems \cite{gr76a}.

Let us consider an elementary mechanical emitter ($m$-emitter) and an
elementary electromagnetic emitter ($e$-emitter). By the elementary we mean
a source that occupies a volume of order $\lambda_s^3$ in the first
case and $\lambda_e^3$ in the second case; $\lambda_s$ and $\lambda_e$
are the wavelengths of the acoustic and electromagnetic waves. A vibrating
object and an oscillating electromagnetic wave in a cavity are examples of
$m-$ and $e-$ emitters respectively. For the comparison to be fair, the
sources are assumed to emit g.w. with one and the same wavelength $\lambda_g$
and are placed at the same distance $R_0$ from the observer.     
 
Let $A$ be the amplitude of elastic vibrations in the $m$-emitter.
Then the characteristic amplitude of the stress tensor $T_{ij}$ is
$\sigma_m \approx \rho_m v_s^2 (A/\lambda_s)$, where $\rho_m$ is the density
of the material and $v_s^2$ is the square of the sound speed. Since
$\nu_g = c/\lambda_g \approx v_s/\lambda_s$, we have $\lambda_s \approx
(v_s/c) \lambda_g \ll \lambda_g$ which means that the body of the 
elementary $m$-emitter is situated deeply inside the inductive zone of
the gravitational radiation and, hence, the retardation effects 
within the source are negligibly small. For the
g.w. amplitude we obtain from Eq. (\ref{ampl2}): 
\be
\label{hm}
h_m \approx \frac{G}{c^4} \frac {1}{R_0} \sigma_m \lambda_s^3.
\en

Let us now consider an elementary $e$-emitter. The amplitude $\sigma_e$
of the electromagnetic stress has the order of magnitude of the energy
density ${\epsilon}_e$ of the varying electromagnetic field, 
$\rho_e c^2 ={\epsilon}_e$, 
i.e.  $\sigma_e \approx {\epsilon}_e$. Since $\nu_g = c/\lambda_g \approx
c/\lambda_e$, the volume of the elementary $e$-emitter is of the
order of $\lambda_g^3$, that is, it is still at the limit of 
applicability of Eq. (\ref{ampl2}) with the retardation effects 
ignored. The amplitude of the emitted gravitational wave is
\be
\label{he}
h_e \approx \frac{G}{c^4} \frac {1}{R_0} \sigma_e \lambda_g^3.
\en

The ratio of (\ref{hm}) and (\ref{he}) is
\be
\label{ratio}
\frac{h_m}{h_e} \approx \frac{\rho_m}{\rho_e} \left( \frac{v_s}{c} \right)^5
\frac{A}{\lambda_s}.
\en
Using the reasonable parameters: $\rho_m \approx 1 g/cm^3, ~ \rho_e
\approx 10^{-18} g/cm^3, ~ v_s/c \approx 10^{-5}, ~ A/\lambda_s \approx
10^{-3}$, we find that $h_m/h_e \approx 10^{-10}$. In other words,
an elementary $e$-emitter is much more efficient than an elementary
$m$-emitter. However, the comparison is not entirely fair as the 
volume of the former, $\lambda_g^3$, is much larger than the volume 
of the later, $\lambda_s^3$. In the volume of an elementary $e$-emitter
one can accomodate a large number $N = (\lambda_g/\lambda_s)^3 \approx
(c/v_s)^3 \gg 1$ of elementary $m$-emitters. Under the condition that they
all are phased to work coherently, the total g.w. amplitude is the sum of
individual amplitudes, and it can be
as large as $h_{mc} \approx N h_m$. Then, we finally obtain 
\be
\label{ratio2}
\frac{h_{mc}}{h_e} \approx \frac{\sigma_m}{\sigma_e}.
\en

This formula answers all the principal questions. We see that the most 
important parameter is the maximal reachable amplitude of the dynamical 
stresses. Extremely high stresses, at the limit of static breaking point, 
for very strong materials, is in the range of 
$\sigma_m \approx 10^{9} dyne~cm^2$. Similar variable 
stresses can be obtained by producing and maintaining
variable electromagnetic fields with characteristic field strength 
$E \approx H \approx 10^{5} gauss$. Then, an elementary $e$-emitter, 
having the volume $\lambda_g^3 \approx 10^{6} cm^3$, produces a g.w. 
amplitude $h \approx 10^{-36}$ at the boundary of the wave zone,
and emits the total power $W \approx 4\times 10^{-13} erg/sec$. To raise 
the amplitude and power, one would have to implement large composite systems.

This analysis illustrates the advantages of the electromagnetic systems.
First, the coherence of the source is automatically achieved in a 
large volume $\sim \lambda_g^3$. The huge number of mechanical
emitters placed in the same volume would need to be specially phased in
order to achieve the constructive interference of the emitted 
gravitational radiation. Second, it seems that it is much
easier to manufacture a simple electromagnetic emitter (essentially,
an oscillating eigen-mode of the electromagnetic  field in a cavity) 
than try to pack together a large
number of specially phased mechanical emitters. Third, the efficient 
generation process suggests the similar (inverse) process of detection. 
It seems natural to use the electromagnetic systems also as the 
detection technique.

\section{Complete laboratory experiment}

The important concern for the succcess of the laboratory experiment is to 
ensure the focusing, as much as possible, of the emitted g.w. power in 
one place, rather than to let it be 
dispersed over all directions. This was one of motivations for the specific
geometrical configuration suggested in \cite{grsazh}. It is proposed
that the emitter is a torus-like electromagnetic resonator with a rectangular 
cross-section (see Fig. 1). The oscillating electromagnetic field in the 
resonator produces a standing gravitational wave in the focal region, near 
the axis of symmetry. The gravitational wave is standing because
the emitted cylindrical gravitational wave passes through the axis of
symmetry and interferes with itself. The
gravitational-wave frequency $\Omega$ is twice the frequency of the variable
electromagnetic field in the generator, $\Omega = 2\omega$. Another resonator 
is placed in the focal region and plays the role of the detector. Its resonant
frequency is tuned to the frequency of the gravitational wave $\Omega$. 
Among the advantages of this particular configuration is its geometrical
simplicity, which allows one to find exact solutions to the  Maxwell 
equations, both, in the generator and in the detector.

\begin{figure}[tbh]
\centerline{\epsfxsize=10cm \epsfbox{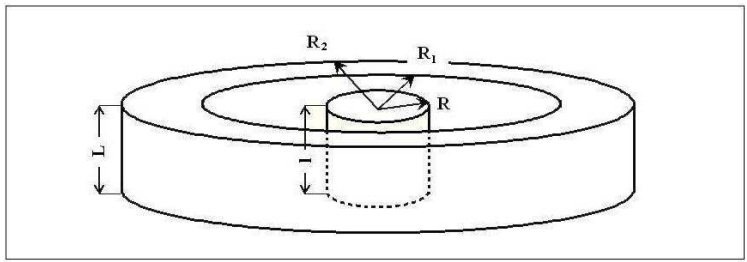}}
\caption{Scheme of a possible laboratory experiment}
\label{Fig.1}
\end{figure}

Although almost all the calculations \cite{grsazh} have been performed exactly,
it is convenient to use the order of magnitude evaluations which follow 
from the calculations, and which, of course, coincide with the
evaluations outlined above. The very important estimate, which we
did not discuss yet, is the detectabilty condition.  

The analysis shows that the change of energy $\Delta {\cal E}$ in the 
resonator-detector depends on whether the initial electromagnetic field
in the detector is realised as a constant (say, magnetic) field, or as a 
an oscillating resonant electromagnetic eigen-mode. It is assumed that
the accumulation of the signal is taking place during the relaxation
time $\tau^{*} =Q/\Omega$, where $Q$ is the quality factor of the 
resonator-detector. Then, in the first case, $\Delta {\cal E} \approx
(h Q)^2 {\cal E}$, where ${\cal E} = (H^2/8 \pi) V$. And in the second
case,  $\Delta {\cal E} \approx (h Q){\cal E}$, 
where ${\cal E} = (E^2/4 \pi) V$. 
These formulas show that the responses of the detector are different
in these two options. But the natural electromagnetic
noises are also different. One should impose different criteria in order to
determine whether the one and the same g.w. signal is detectable or not. 
In the first case,
one can think that one can distinguish one ``photon" (a resonant mode
exitation with energy ${\cal E} \sim \hbar \Omega$) on the background of
a constant (magnetic) field. In the second case, one should typically assume
that only the number $\sqrt N$ of new ``photons" can be distinguished on the 
background of the already existing $N$ ``photons", 
$N = {\cal E}/ \hbar \Omega$.
Combining each of responses with the corresponding detectability condition,
one finds that there is in fact no much difference between these two cases
in terms of the measurable amplitude $h$. The order of magnitude evaluation
shows that the detectable $h$ is
\be
\label{det}
h_{det} \approx \sqrt{\frac{ \hbar \Omega}{{\cal E}}} \frac{1}{Q}.
\en  

The situation can, however, change considerably in favor of the 
second option, if one succeeds 
in realising a special (quantum) state of the electromagnetic field in the 
detector, such that the variance of the number of quanta $N$ is much smaller 
than $\sqrt N$. Then, assuming that the quantum non-demolition measurement
can distinguish the energy of, say, a few new quanta on the backgound of a 
huge (mean) number of quanta in the detector, the detectable g.w. 
amplitude $h_{det}$ can be significantly
lowered. We will first continue our estimates without taking this 
possibility into account, but will return to it later.

The geometry of the system allows one to choose the following optimal 
parameters: $L \approx l \approx \lambda$, 
where $\lambda = 2\pi c/\Omega$, $R \approx 2 \lambda/3,
~ R_1 \approx 2 \lambda, ~ R_2 \approx 7\lambda/2$. Then the detectability
condition can be written as 
\be
\label{lab}
\lambda^4 E^2 H Q \approx 30 \frac{c^4}{G} \sqrt {\hbar c},
\en
where $E$ is the amplitude of the oscillating field in the generator 
and $H$ is the typical (possibly, constant magnetic) field in the 
detector. Let us take for illustration $\lambda \approx 10^2 cm, ~
E \approx H \approx 3 \times 10^{5} gauss, ~ Q \approx 7\times 10^{13}$. 
Then, the
left-hand side of Eq. (\ref{lab}) is 4 orders of magnitude smaller than the
right-hand side of the same formula. It is this gap of 4 orders of 
magnitude that we referred to in the Abstract. 

One possibility to satisfy Eq. (\ref{lab}) is to simply increase the 
size of the system, going from  $\lambda \approx 10^2 cm$ to 
$\lambda \approx 10^3 cm$. Another possibility
would be to raise the product $E^2 HQ$ by the same 4 orders of magnitude. 
Probably the most elegant and realistic possibility is to try and 
improve the detectability condition by
using  the sophisticated quantum states of the electromagnetic field in 
the detector, that we mentioned earlier. Certainly, the laboratory 
experiment is going to be difficult, but it seems worth of trying.

\section{Electromagnetic detectors for relic gravitational waves}

In laboratory conditions, and in almost all astrophysical situations,
one can completely neglect the non-linearities of gravity,  
that is, the interaction of gravitational waves with
other gravitational fields. However, this is not always the case.  
The most important example is the 
interaction of gravitational waves with the strong variable gravitational 
field of the very early Universe. One can use the engineering intuition
in order to understand what is going on here.  A gravitational wave can be 
thought of as a harmonic oscillator, while the smooth variable 
gravitational field of the surrounding Universe as a gravitational 
pump field. The g.w. oscillator is
parametrically coupled to the gravitational pump field. This specific 
coupling follows from the non-linear structure of the Einstein equations. 
This coupling provides a mechanism for the superadiabatic (parametric) 
amplification of classical waves and for the quantum-mechanical generation 
of waves from their zero-point quantum oscillations \cite{gr03}. 
The word ``superadiabatic" emphasizes the fact that this effect takes place 
over and above whetever happens to the wave during very slow
(adiabatic) changes of the pump field. That is, we are interested 
in the increase of occupation numbers, rather than in the gradual 
shift of energy levels. The word ``parametric" emphasizes the mathematical
structure of the wave equation. A parameter of the oscillator,
namely its frequency, is being changed by the variable pump
field. It is this sufficiently rapid change of frequency of the
oscillator that is 
responsible for the considerable increase of energy of that oscillator. 

The parametric amplification of the inevitable zero-point quantum 
oscillations leads to the generation of a stochastic background of 
relic gravitational waves. The mechanism itself is based on the 
fundamental physics only, but the generated signal depends on the 
pump field. In other words, the amount of relic gravitational
waves depend on a concrete cosmological model of
the very early Universe. Combining the theory with the available 
cosmological data, one
can evaluate the expected level of high-frequency relic gravitational 
waves \cite{gr03}. For example, the root-mean-square (r.m.s.) amplitude
is expected to reach $h_{r.m.s.} \approx 10^{-30}$ at $\nu=10^{7} Hz$. 
The amplitude is smaller at higher frequencies. It may reach
$h_{r.m.s.} \approx 10^{-32}$ at $\nu =10^{11} Hz$, but then it 
should quickly decrease as a function of higher frequencies. There
is no much sense to expect any relic gravitational waves at frequencies
higher than that.
 
We may be lucky (although it does not seem very likely) if the 
thermal background of gravitational waves 
survived until now. Then, in the vicinity of $\nu =10^{11} Hz$  
there will be a maximum of the Planck spectrum. The amplitude
of the Planck spectrum can be in the region of $h_{r.m.s.} \approx
3 \times 10^{-32}$, but not very much larger than this. The quoted numbers
for relic and thermal backgrounds give the feeling of what we can 
expect of high-frequency cosmic gravitational waves.

What can be said about the detectability of relic gravitational waves ?
The first impression is that they may be easier to detect than
the gravitational waves produced in laboratory. For example,
at $\nu = 10^{7} Hz$, the amplitude of relic gravitational waves,
$h_{r.m.s.} \approx 10^{-30}$, is several orders of magnitude higher 
than the realistic amplitude of laboratory gravitational
waves (see Sec. IV). Unfortunately, this does not mean that the
relic gravitational waves are easier to detect. The crucial difference 
is in the character of these two signals. The laboratory signal is
essentially a deterministic monochromatic wave, which allows one to 
systematically accumulate the response amplitude. In contrast, relic
gravitational waves form a random signal, which allows only an 
accumulation of energy. When it comes to evaluation of the detectable 
amplitude of relic gravitational waves \cite{gr76b}, 
formula (\ref{det}) should be modified. Under the same conditions that
formula (\ref{det}) was derived, but now for the stochastic signal, we 
obtain
\be
\label{det2}
h_{det} \approx \sqrt{\frac{ \hbar \Omega}{{\cal E}}} \frac{1}{\sqrt{Q}}.
\en  
Applying this formula to a simple electromagnetic detector, we can 
expect to measure $h_{det} \approx 10^{-26}$ instead of the required
level of the cosmic signal $h_{r.m.s.} \approx 10^{-30}$. 

This gap is sufficiently wide to discount hopes to bridge this gap by
longer observation time, or cross-correlation of several detectors, etc. 
It appears that the sensitivity of electromagnetic detectors 
can reach the required level only if the squeezed quantum states of the 
electromagnetic field in the detector, and quantum non-demolition 
measurements, are implemented.  Aternatively, one can think of large 
composite systems (``large crystal" \cite{gr76b}), where the
individual elements are arranged to work more coherently than simply
absorbing gravitational radiation independently one from other. 
Although these techniques seem possible in principle, so far, there is 
no concrete proposals how to implement them in practice.
Regrettably, the detection of high-frequency relic gravitational 
waves will probably be even more challenging problem than 
their detection in low-frequency and very low-frequency bands.
Smart ideas are badly needed.

\end{document}